\documentclass[preprint2]{aastex}
\usepackage{amsmath}
\usepackage[varg]{txfonts}
\usepackage{sistyle}
\usepackage{graphicx}
\usepackage{textcomp}
\bibpunct{(}{)}{;}{a}{}{,} 
\usepackage{color}

\renewcommand{\eqref}[1]{Eq.~\ref{#1}}
\newcommand{\fref}[1]{Figure~\ref{#1}}
\newcommand{\ie}{i.\,e.}

\newcommand{\eg}{e.\,g.}

\title{What asteroseismology can do for exoplanets:\\
Kepler-410A b is a Small Neptune around a bright star, in an eccentric orbit consistent with low obliquity}
 
\author{V.~Van~Eylen$^{1,2\star}$, M.~N.~Lund$^{1,4}$, V.~Silva~Aguirre$^{1}$, T.~Arentoft$^{1}$, H.~Kjeldsen$^{1}$, S.~Albrecht$^{3}$, W.~J.~Chaplin$^{5}$, H.~Isaacson$^{6}$, M.~G.~Pedersen$^{1}$, J.~Jessen-Hansen$^{1}$, B.~Tingley$^{1}$, J.~Christensen-Dalsgaard$^{1}$, C.~Aerts$^{2}$, T.~L.~Campante$^{5}$ and S.~T.~Bryson$^{7}$}

\affil{$^1$ Stellar Astrophysics Centre, Department of Physics and Astronomy, Aarhus University, Ny Munkegade 120, \\
DK-8000 Aarhus C, Denmark}
\affil{$^2$ Instituut voor Sterrenkunde, Katholieke Universiteit Leuven, Celestijnenlaan 200 B, B-3001 Heverlee, Belgium\label{inst2}}
\affil{$^3$ Department of Physics, and Kavli Institute for Astrophysics and Space Research, Massachusetts Institute of Technology, Cambridge, MA 02139, USA\label{inst3}}
\affil{$^4$ Sydney Institute for Astronomy (SIfA), School of Physics, University of Sydney, NSW 2006, Australia\label{inst4}}
\affil{$^5$ School of Physics and Astronomy, University of Birmingham, Edgbaston, Birmingham, B15 2TT, UK\label{inst5}}
\affil{$^6$ Department of Astronomy, University of California, Berkeley, CA 94820, USA\label{inst6}}
\affil{$^7$ NASA Ames Research Center, Moffett Field, CA 94035}

\email{$^{\star}$vincent@phys.au.dk}

\shorttitle{A study of Kepler-410}
\shortauthors{Van Eylen et al.}

\received{receipt date}
\revised{revision date}
\accepted{acceptance date}

\begin{abstract}
We confirm the \textit{Kepler} planet candidate Kepler-410b (KOI-42b) as a Neptune sized exoplanet on a $17.8$~day, eccentric orbit around the bright ($K_\textrm{p} = 9.4$) star Kepler-410A. This is the third brightest confirmed planet host star in the \textit{Kepler} field and one of the brightest hosts of all currently known transiting exoplanets. Kepler-410 consists of a blend between the fast rotating planet host star (Kepler-410A) and a fainter star (Kepler-410B), which has complicated the confirmation of the planetary candidate. Employing asteroseismology, using constraints from the transit light curve, adaptive optics and speckle images, and \textit{Spitzer} transit observations, we demonstrate that the candidate can only be an exoplanet orbiting Kepler-410A. Via asteroseismology we determine the following stellar and planetary parameters with high precision; M$_\star = 1.214 \pm 0.033$ M$_\odot$, R$_\star = 1.352 \pm 0.010$ R$_\odot$, Age $= 2.76 \pm 0.54$ Gyr, planetary radius ($2.838 \pm 0.054$ R$_\oplus$), and orbital 
eccentricity ($0.17^{+0.07}_{-0.06}$). In addition, rotational splitting of the pulsation modes allows for a measurement of Kepler-410A's inclination and rotation rate. Our measurement 
of an inclination of $82.5^{+7.5}_{-2.5}$ [$^\circ$] indicates a low obliquity in this system. Transit timing variations indicate the presence of at least one additional (non-transiting) planet in the system.
\end{abstract}


\begin{document}

\maketitle

\section{Introduction}

Launched March 2009, the \textit{Kepler} mission continuously observed a field in the sky centered on the Cygnus-Lyra region with the primary goal of detecting (small) exoplanets, by photometrically measuring planetary transits to a high level of precision \citep{borucki2008}. Apart from a growing list of confirmed planets (currently 152), the \textit{Kepler} catalog contains 3548 planetary candidates \citep{batalha2013}. The order of magnitude difference between those numbers illustrates the intrinsic difficulty of exoplanet confirmation. 

Stars showing transit-like features are termed Kepler Objects of Interest (KOIs). Here we study KOI-42 (KIC 8866102, HD 175289, subsequently refered to as Kepler-410), which shows transit-like features consistent with a small planet ($R_\textrm{p} \approx 2.6 $ R$_\oplus$) on a relatively long orbit \citep[17.83 d;][]{borucki2011}. Apart from the bright host star (\textit{Kepler} magnitude $K_\textrm{p}$ = 9.4) Kepler-410 also consists of a fainter blended object \citep[$K_\textrm{p}$ = 12.2, ][]{adams2012}. We refer to this object as Kepler-410B, while we use Kepler-410A for the bright host star. The brightness of the system would make it a prime target for follow-up studies, if it can be confirmed that the transits are indeed occurring around Kepler-410A. Unfortunately the added complexity due to the presence of Kepler-410B, and the presumably small mass of the planet candidate, has so far prevented the planetary candidate to be confirmed as planet, or shown to be a false positive.

In this paper we will show that the transit-like features are indeed caused by a planet orbiting Kepler-410A. For this we combine information from the well-determined transit shape with additional (ground-based) observations and \textit{Spitzer} measurements. We also take advantage of Kepler-410 being almost exclusively observed in \textit{Kepler}'s short-cadence mode \citep[sampling it every 58.8 s,][]{borucki2008}, which allows for the detection of solar-like oscillations. Analyzing the stellar pulsations aids the confirmation of Kepler-410A as planet host and leads to accurate determination of the stellar parameters. We further measure the stellar rotation and its inclination by analyzing the pulsation modes. Such an analysis was recently carried out for Kepler-50 and Kepler-65 by \citet{2013ApJ...766..101C}.

In \S~\ref{sec:asteroseismology}, we describe the asteroseismic modeling before we present the various arguments that validate Kepler-410A b as a planet in \S~\ref{sec:planetary_validation}. The planetary and orbital parameters are presented in \S~\ref{sec:planetary_parameters}. We discuss the characteristics of the the system in \S~\ref{sec:discussion} and our conclusions are presented in \S~\ref{sec:conclusion}.



\section{Stellar properties from asteroseismology}
\label{sec:asteroseismology}
\begin{figure*}[ht]
\centering
\includegraphics[scale=0.48]{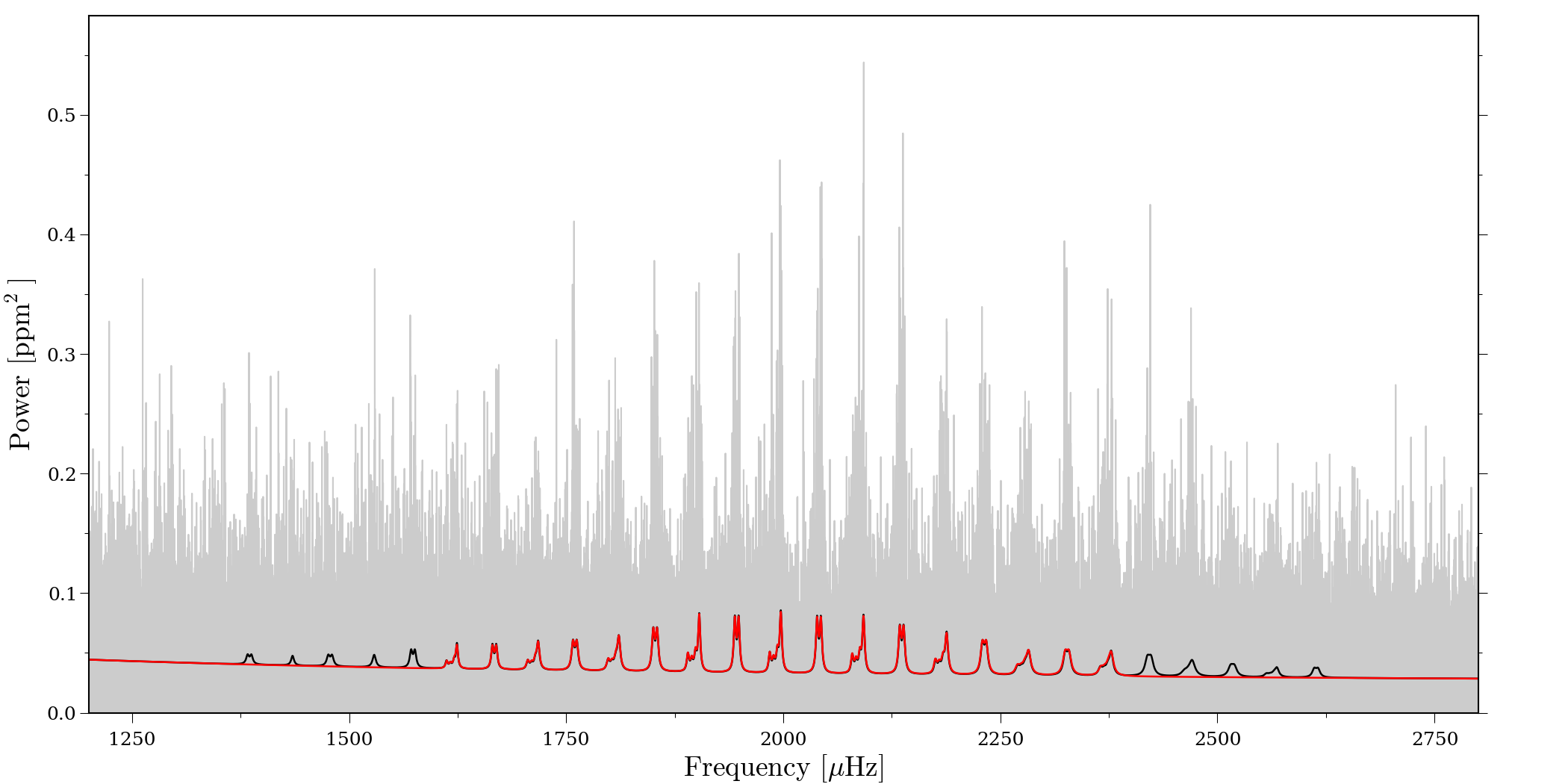}
\caption{ \emph{ \footnotesize Power spectrum of Kepler-410 (gray). Overlain are the model fits (\eqref{eq:limitspec}) obtained from the MCMC peak-bagging. The black curve gives the model when including modes from the range $1370-2630\, \rm \mu Hz$ - all mode frequencies in this range were included in the stellar modeling. The red curve gives the model obtained when excluding the five outermost modes obtained in the first fit (black curve) in each end of the frequency scale. From this fit we get the estimates of the stellar inclination and frequency splitting.}}
\label{fig:power_spectrum}
\end{figure*}

Kepler-410 was observed in short-cadence mode for the entire duration of the \textit{Kepler} mission, except during the second quarter of observations (Q2) where the long cadence mode was used. The latter observations are not included in the asteroseismic analysis, and we use short-cadence simple aperture photometry (SAP) data from Q0-Q1 and Q3-Q13. Before using the data as input for asteroseismology, it is de-trended and normalized using a specifically designed median filter to remove all transit features from the time series. The resulting time series is then used to derive a power spectrum\footnote{The power spectrum was calculated using a sine-wave fitting method \citep[see, \eg,][]{1992PhDT.......208K,1995A&A...301..123F} which is normalized according to the amplitude-scaled version of Parseval's theorem \citep[see, \eg,][]{1992PASP..104..413K}, in which a sine wave of peak amplitude, A, will have a corresponding peak in the power spectrum of $\rm A^2$.}, which is shown in \fref{fig:power_spectrum}.

\subsection{Asteroseismic frequency analysis}
The extraction of mode parameters for the asteroseismic analysis was performed by \emph{Peak-bagging} the power spectrum \citep[see, \eg,][]{2003Ap&SS.284..109A}. This was done by making a global optimization of the power spectrum using an \emph{Markov Chain Monte Carlo} (MCMC) routine\footnote{The program StellarMC was used, which was written and is maintained by Rasmus Handberg.}, including a parallel tempering scheme to better search the full parameter space \citep[see][]{2011A&A...527A..56H}. In the fit the following model was used for the power spectrum:

\begin{equation}
\centering
\mathcal{P}(\nu_j ; \mathbf \Theta) = \sum_{n=n_{a}}^{n_{b}}\sum_{\ell=0}^{2}\sum_{m=-\ell}^{\ell} \frac{\mathcal{E}_{\ell m}(i) \tilde{V}_{\ell}^2 \alpha_{n\ell}}{1+\frac{4}{\Gamma_{n\ell}^2}(\nu-\nu_{n\ell m})^2 } + B(\nu), 
\label{eq:limitspec}
\end{equation}
here $n_a$ and $n_b$ represent respectively the first and last radial order included from the power spectrum. We include modes of degree $\ell=0-2$. Each mode is described by a Lorentzian profile \citep[see, \eg,][]{1990ApJ...364..699A, 2003ApJ...589.1009G} due to the way in which the p-modes are excited, namely stochastically by the turbulent convection in the outer envelope upon which they are intrinsically damped \citep{1994ApJ...424..466G}. In this description $\nu_{n\ell m}$ is the frequency of the mode while $\Gamma_{n\ell}$ is a measure for the damping\footnote{The mode life time is given by $\tau=1/\pi\Gamma_{n\ell}$.} rate of the mode and equals the full width at half maximum of the Lorentzian. 
$\mathcal{E}_{\ell m}(i)$ is a function that sets the relative heights between the azimuthal $m$-components in a split multiplet as a function of the stellar inclination \citep[see, \eg,][]{1977AcA....27..203D, 2003ApJ...589.1009G}. The factor $\tilde{V}_{\ell}^2$ is the relative visibility (in power) of a mode relative to the radial and non-split $\ell=0$ modes. The factor $\alpha_{n\ell}$ represents an amplitude modulation which mainly depends on frequency and is generally well approximated by a Gaussian. 

In this work we do not fix the relative visibilities, as recent studies \citep[see, \eg,][]{2010A&A...515A..87D,2011A&A...528A..25S,lund_to_come} have suggested that the theoretical computed values \citep[see, \eg,][]{2011A&A...531A.124B} are generally not in good agreement with observations. In line with this notion we find that the theoretically expected values for the relative visibilities from \citet[][]{2011A&A...531A.124B}, which, using the spectroscopic parameters for Kepler-410 (see Table~\ref{table:spectroscopic_parameters2}), are given by $\tilde{V}_{\ell=1}^2\approx 1.51$ and $\tilde{V}_{\ell=2}^2\approx 0.53$, do not conform with the values obtained from our optimization (see Table~\ref{table:final_parameters}).

We describe the granulation background signal given by $B(\nu)$ by a sum of powerlaws \citep{1985ESASP.235..199H}, specifically in the version proposed by \cite{KarPhD}:
\begin{equation}
\centering
B(\nu) = B_n +\sum_{i=1}^{2}{\frac{4\sigma_i^2 \tau_i}{1 + (2\pi \nu \tau_i)^2 + (2\pi \nu \tau_i)^4}}.
\label{eq:bg}
\end{equation}
In this equation $\sigma_i$ and $\tau_i$ gives, respectively, the rms variation in the time domain and the characteristic time scale for the granulation and the faculae components. The constant $B_n$ is a measure for the photon shot-noise. 

The frequencies of the individual modes in the interval $1370-2630\, \rm \mu Hz$ found from this optimization are used in the stellar modeling, see \S~\ref{sec:stellar_param} and \fref{fig:power_spectrum}.


\subsubsection{Stellar inclination and rotational splitting}
\label{sec:splitting}

Asteroseismology can via a fit of \eqref{eq:limitspec} be used to infer parameters such as the stellar rotation period and inclination\footnote{Going from $i=0^{\circ}$ at a pole-on view to $i=90^{\circ}$ for equator-on view.}. The information on these properties are found from the rotationally induced splitting of a oscillation mode of degree $\ell$ into $2\ell+1$ azimuthal $m$-components with values going from $m=-\ell$ to $m=\ell$.
In the case of a slow stellar rotation the star is generally assumed to rotate as a rigid body and the modes will be split as \citep{1951ApJ...114..373L}:
\begin{equation}
\centering
\nu_{n\ell m} = \nu_{n\ell} + m\frac{\Omega}{2\pi}(1-C_{n\ell}) \approx \nu_{n\ell} + m\nu_s,
\label{eq:split}
\end{equation}
with $\nu_{n\ell m}$ being the frequency entering into \eqref{eq:limitspec}, while $\nu_{n\ell}$ gives the unperturbed resonance frequency. The azimuthal order of the mode is given by $m$, $\Omega$ is the angular rotation rate of the star and $C_{n\ell}$ is the so-called \emph{Ledoux constant}; a dimensionless quantity describing the effect of the Coriolis force. For high-order, low-degree solar oscillations, as the ones seen in Kepler-410A, this quantity is of the order $C_{n\ell} < 10^{-2}$ and is therefore neglected. The splittings can thereby be seen as being dominated by advection. In this way we see that the splitting due to rotation between adjacent components of a multiplet will approximately be $\nu_s = \Omega/2\pi$, which will be referred to as the rotational frequency splitting.

In the optimization we use for the inclination a flat prior in the range $0-180^{\circ}$ and then fold the results from the MCMC around $i=90^{\circ}$. The reason for this is to better sample the posterior of the inclination very close to $i=90^{\circ}$, which would not be possible with a fixed boundary for the inclination at $i=90^{\circ}$ \citep[see, \eg,][]{2013ApJ...766..101C}. A correct sampling of this region mainly has an influence on the credible regions computed for the value of the inclination. For the splitting we use a flat prior in the range $0-5\, \rm \mu Hz$. 
 
\begin{figure*}[ht]
\centering
\includegraphics[scale=0.5]{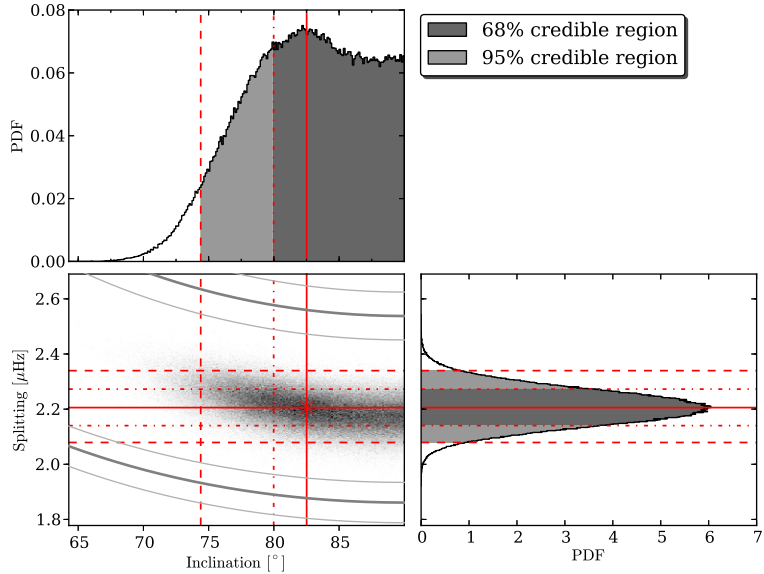}
\caption{ \emph{ \footnotesize Rotational splitting and inclination angle for Kepler-410 from the MCMC peak-bagging. The bottom left panel shows the correlation map between the inclination and the rotational splitting, while the panels above (inclination) and to the right (splitting) give the  marginal probability density (PDF) functions for these two parameters. Our estimates for the parameters are given by the median values of their respective PDFs, indicated by the solid lines. The $68\%$ credible regions (found as the highest posterior density credible regions) are indicated by the dark gray part of the PDFs (bounded by dash-dotted lines), while the light gray indicates the additional part of the PDFs covered in a $95\%$ credible regions (bounded by dashed lines). The PDF for the inclination was found after first having folded the part of the full distribution ($0-180^{\circ}$) in the range $90-180^{\circ}$ onto the part in the range $0-90^{\circ}$. In the correlation map we have indicated the splitting as a 
function of inclination (dark gray), with associated uncertainty (light gray), corresponding to the values 
of $v\sin(i)$ estimated by \citet{molenda2013} (bottom lines) and \citet[][]{2013ApJ...767..127H} (top lines) and the radius estimate from our analysis.}}
\label{fig:split}
\end{figure*}

For the estimation of the inclination and rotational splitting we did not include the entire range used in estimating frequencies for the modeling, see \fref{fig:power_spectrum}. The rationale for using a narrower range that excludes the modes at highest and lowest frequencies is that we want only the modes with the highest signal-to-noise ratio. Furthermore, for modes at high frequencies the mode width becomes problematic for a proper estimate of the splitting.

Figure~\ref{fig:split} shows the correlation map from the MCMC analysis for the stellar inclination ($i$) and rotational splitting ($\nu_s$), going from low (light) to high (dark). The adopted values for the inclination and splitting are found by the median of the marginalized distribution, and indicated in the figure by the intersection of the two solid lines, final values are given in Table~\ref{table:final_parameters}. The dark gray part of the PDFs (bounded by dash-dotted lines) show the $68\%$ highest posterior density credible regions for each parameter and serve as the error for our estimates. The light gray (dashed lines) indicate the additional part of the PDFs covered by the $95\%$ credible regions. We find a value of $i=82.5^{\circ +7.5}_{-2.5} $ for the stellar inclination of the star, indicating a nearly equator-on view.

\begin{figure}[ht]
\centering
\resizebox{\hsize}{!}{\includegraphics{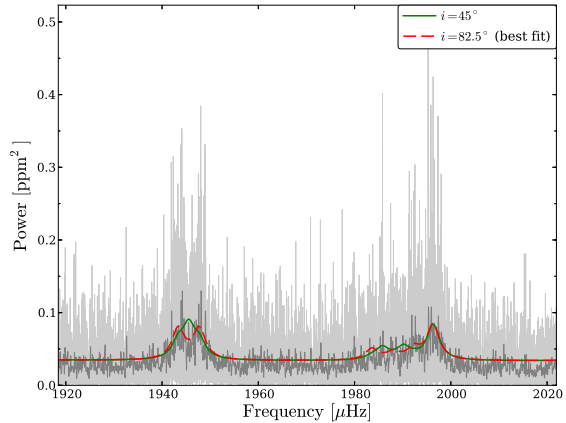}}
\caption{ \emph{ \footnotesize Power spectrum of Kepler-410 (light gray) for a single order, with the dark gray giving the $0.1\, \rm \mu Hz$ smoothed version. Overlain is the model fit obtained from the MCMC peak-bagging (dashed red), in addition to the limit spectrum obtained when using an inclination angle of $45^{\circ}$ (full green).}}
\label{fig:single_order}
\end{figure}

Figure~\ref{fig:single_order} shows, for a single order, the best fitting model in red (\ie\ using $i=82.5^{\circ}$). In green the limit spectrum is given when instead using a value for the stellar inclination of $i=45^{\circ}$ and keeping all other parameters fixed to the best fit values. This shows the large effect of the stellar inclination on the appearance of the limit spectrum from the variation in the relative heights of different azimuthal components.

To compare our result with the literature, we can compute the value of the splitting from literature values of $v\sin(i)$ via:
\begin{equation}
\nu_s = \frac{ [ v\sin(i) ] }{2\pi R \sin(i)}.
\label{eq:vsini}
\end{equation}
Using the radius found from the asteroseismic modeling (see \S~\ref{sec:stellar_param}), we have in \fref{fig:split} illustrated the corresponding values for the splitting from the estimate of $v\sin(i)$ by \citet{molenda2013} of $11.0\pm0.8\,\rm kms^{-1}$, and \citet[][]{2013ApJ...767..127H} of $15.0 \pm 0.5\,\rm kms^{-1}$. From our asteroseismic modeling we get, as expected from \fref{fig:split}, a value in between these estimates of $v\sin(i) = 12.9 \pm 0.6\,\rm kms^{-1}$ (see Table~\ref{table:final_parameters}).


\subsection{Asteroseismic modeling}
\label{sec:stellar_param}

\begin{table*}[ht]
\caption{ \emph{ \footnotesize Stellar parameters from spectroscopy. \textit{a:} \citet[][]{2013ApJ...767..127H}. \textit{b:} \citet{molenda2013}.}}
\label{table:spectroscopic_parameters2}
\centering
\begin{tabular}{l l l l l l}
\hline\hline\\[-0.35cm]
  Reference &  $\rm T_{eff}$ (K) &  $\log\,g$ & $\rm [Fe/H]$ (dex) & $ v\rm\sin(i)$ ($\rm km s^{-1}$) & Instrument \\[0.05cm]
\hline\\[-0.3cm]
 \textit{a}	&  $6325 \pm 75$  &  -  				& $+0.01\pm 0.10$ 	& $15.0\pm0.5$ 	& HiRES, McDonald \\
 \textit{b}	&  $6195\pm 134$  &  $3.95\pm 0.21$ 	& $-0.16 \pm 0.21$ 	& $11.0\pm 0.8$ 	& ESPaDOnS 			\\[0.05cm]
\hline
\end{tabular}
\end{table*} 

The stellar parameters were determined based on grids of models constructed using the GARching STellar Evolution Code \citep[GARSTEC,][]{Weiss:2008jy}. The input physics consists of the NACRE compilation of nuclear reaction rates \citep{Angulo:1999kp}, the \citet{Grevesse:1998cy} solar mixture, OPAL opacities \citep{Iglesias:1996dp} for high temperatures complemented by low-temperature opacities from \citet{Ferguson:2005gn}, the 2005 version of the OPAL equation of state \citep{Rogers:1996iv}, and the mixing-length theory of convection as described in \citet{2013sse..book.....K}. One grid of models also included the effect of convective overshooting from the stellar core when present. This is implemented in GARSTEC as an exponential decay of the convective velocities in the radiative region, and the used efficiency of mixing is the one calibrated to reproduce the CMD of open clusters \citep[\eg][]{magic2010}. Diffusion of helium and heavy elements was not considered.

Our grid of models spans a mass range between 1.10-1.40~M$_\odot$ in steps of 0.02~M$_\odot$, and comprises five different compositions for each mass value spanning the 1-$\sigma$ uncertainty in metallicity as found from spectroscopy by \citet[][]{2013ApJ...767..127H}, see Table~\ref{table:spectroscopic_parameters2}. We chose this set of atmospheric constraints for the host star since they were derived using an asteroseismic determination of the surface gravity to avoid degeneracies from the correlations between $T_{\rm eff}$, $\log\,g$, and [Fe/H] \citep[see][for a thorough discussion]{2012ApJ...757..161T}. While the relative abundance of heavy elements over hydrogen can be directly determined from the measured [Fe/H] value, the assumption of a galactic chemical evolution law of $\Delta Y/\Delta Z = 1.4$ \citep[\eg,][]{Casagrande:2007ck} allows a complete determination of the chemical composition. For both grids of models we computed frequencies of oscillations using the Aarhus Adiabatic Oscillations 
Package \citep[ADIPLS,][]{ChristensenDalsgaard:2008kr}, and determined the goodness of fit by calculating a $\chi^2$ fit to the spectroscopic data and frequency combinations sensitive to the interior as described in \citet{SilvaAguirre:2013in}. Final parameters and 
uncertainties were obtained by a 
weighted mean and standard deviation using the $\chi^2$ values 
of the grid without overshooting, and we added in quadrature the difference between these central values and those from the grid with overshooting to encompass in our error bar determinations the systematics introduced by the different input physics.

By combining the \citet{Casagrande:2010hj} implementation of the InfraRed Flux Method (IRFM) with the asteroseismic determinations as described in \citet{SilvaAguirre:2011es,SilvaAguirre:2012du}, it is possible to obtain a distance to the host star which is in principle accurate to a level of ${\sim}5\%$. Since the photometry of the host star might be contaminated by the close companion, we carefully checked the 2MASS photometry used in the implementation of the IRFM for warnings in the quality flags. The effective temperature determined by this method of $T_{\rm eff}=6273\pm140$K is in excellent agreement with those given in Table~\ref{table:spectroscopic_parameters2}, giving us confidence that the distance to the host star is accurately determined.

The final parameters of the star, including this distance, are given in Table~\ref{table:final_parameters}. From the stellar model parameters obtained from the peak-bagged frequencies we can calculate the Keplerian (rotational) break-up frequency of the star as:
\begin{equation}
\frac{\Omega_{\rm K}}{2\pi} = \rm \frac{1}{2\pi}\sqrt{\frac{GM}{R^3}} \approx 70.0 \pm 1.4\, \mu \rm Hz,  
\end{equation}
whereby the star rotates at a rate of ${\sim}3\%$ of break-up as $2\pi\nu_s \approx 0.03 \Omega_{\rm K}$. 
With this splitting it is worth considering the effect of second-order perturbations, $\delta\nu_2$, on the rotational frequency splitting:
\begin{equation}
\nu_{n\ell m} = \nu_{n\ell} + m\nu_s + \delta\nu_2.
\end{equation}
As described in Appendix~\ref{append}, this effect produces a small offset in the frequencies of the pulsations that in turn affects the stellar parameters derived from asteroseismic modeling. For this reason we iterated the frequency extraction with the stellar properties until the value of the break-up frequency converged (obtained after only a few iterations). The final model parameters are given in Table.~\ref{table:final_parameters}. We note that the change in parameters from including second-order effects is quite negligible, generally less than one per mil, with the exception of the age which is changed by about ${\sim}1.5\%$.

\begin{figure}[ht]
\centering
\resizebox{\hsize}{!}{\includegraphics{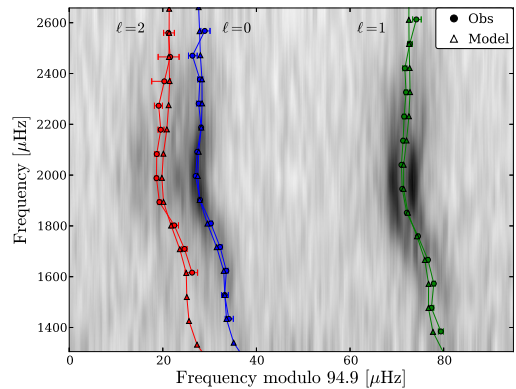}}
\caption{ \emph{ \footnotesize \'Echelle diagram showing, in gray scale, the power spectrum of Kepler-410. Overlaid are the frequencies estimated from the MCMC peak-bagging (circles), along with the frequencies from the best-fitting stellar model after a surface correction (triangles). The frequencies estimated from the peak-bagging are the $m=0$ components, while $|m|>0$ components are included in \eqref{eq:limitspec} by the splitting. For an inclination as found for Kepler-410 mode power will for $\ell=1$ modes mainly be contained in the $m=\pm 1$ components, whereby the estimated $m=0$ component needed for the asteroseismic modeling should as seen be found in-between the $m=\pm 1$ power concentrations (assuming a symmetric splitting). }}
\label{fig:echelle}
\end{figure}

In \fref{fig:echelle} the \'echelle diagram \citep[][]{1983SoPh...82...55G} is shown, with observations overlaid by the frequencies from the best stellar model after the above iteration. For the sake of the comparison in the \'echelle diagram a surface correction has been applied to the model frequencies following the procedure of \citet[][]{2008ApJ...683L.175K}\footnote{As reference frequency we use the mean value of the radial modes, while $b$ is set to the solar calibrated value of $4.823$ \citep{2012ApJ...749..152M}}. Note, that the surface correction is not needed for the model optimization as frequency ratios, unaffected by the surface layers, are used rather than the actual frequencies. The splitting of the $\ell=1$ modes is clearly visible, with mode power mainly contained in the sectoral $m =\pm 1$ azimuthal components around the zonal $m=0$ components found in the peak-bagging and the modeling. This distribution of power between the azimuthal components is a function of the stellar inclination 
angle \citep[see, \eg,][]{2003ApJ...589.1009G}, where we indeed for $i$ close to $90^{\circ}$ (as found for Kepler-410) should expect to see power mainly in the sectoral components of $\ell=1$.

\section{Planetary validation}\label{sec:planetary_validation}

In this Section, we investigate the possible scenarios causing the transit-like features in the \textit{Kepler} data for Kepler-410. In \S~\ref{sec:confirmation_constraints}, we describe the constraints, as provided by the \textit{Kepler} data themselves, \textit{Spitzer} data and additional observations from ground. In \S~\ref{sec:confirmation_scenarios}, we then use those constraints to assess the likelihood of various scenarios, to conclude that the transits are indeed caused by a planet in orbit around Kepler-410A.


\subsection{Constraints\label{sec:confirmation_constraints}}

\label{sec:constraints}

\subsubsection{Geometry of transit signal}
\label{sec:transit_geometry}
\begin{figure}[ht]
\resizebox{\hsize}{!}{\includegraphics{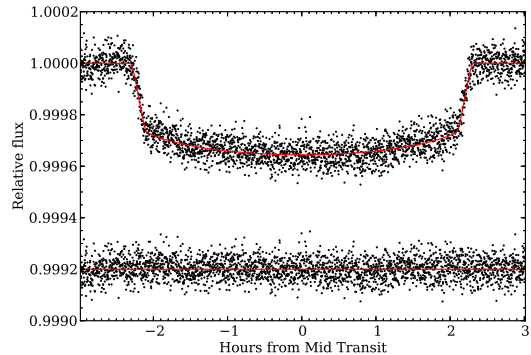}}
\caption{\emph{ \footnotesize Planetary transit using the phase-folded observations (see \S~\ref{sec:period}), which were binned for clarity. The best fit is shown with a red line, together with the residuals (offset).}}
\label{fig:transit_fit}
\end{figure}

A first constraint on what could be causing the transit signal in the \textit{Kepler} data, comes from the \textit{geometry of the transit signal} itself. While the transit signal could be diluted by additional stellar flux (i.e. by Kepler-410B, or additional unseen blends), the shape of the transit, as governed by the four contact points, remains the same. We use $T_\mathrm{tot}$ for the total transit duration, $T_\mathrm{full}$ for the duration between contact points two and three (the transit duration minus ingress and egress), $b$ for the impact parameter, and find \citep[see \eg\ ][]{2010arXiv1001.2010W}:

\begin{equation}
\label{equ:transit_shape}
 \frac{\sin(\pi T_\mathrm{tot} / P)}{\sin(\pi T_\mathrm{full} / P)} =  \frac{\sqrt{(1 + R_\mathrm{p}/R_\star)^2 - b^2}}{\sqrt{(1 - R_\mathrm{p}/R_\star)^2 - b^2}}\, .
\end{equation}
Here $R_\star$ and $R_{\rm p}$ indicate the stellar and planetary radii. This equation (which only strictly holds for a zero-eccentricity orbit) can be understood by considering the most extreme case, namely a binary with two stars of the same size and $b = 0$, causing the equation to go to infinity. The transit becomes fully V-shaped, half of the transit is in ingress, while the other half is in egress. As it turns out, the short-cadence data constrains the transit shape to be clearly different from a V-shape as can be seen in \fref{fig:transit_fit}. With the left-hand side of \eqref{equ:transit_shape} determined by the data and setting $b\equiv0$, an upper limit on $R_\mathrm{p}/R_\star$ can be determined. Given the observed transit depth this ratio can now be used to establish an upper limit on any light dilution.

\begin{figure}[ht]
\centering
\resizebox{\hsize}{!}{\includegraphics{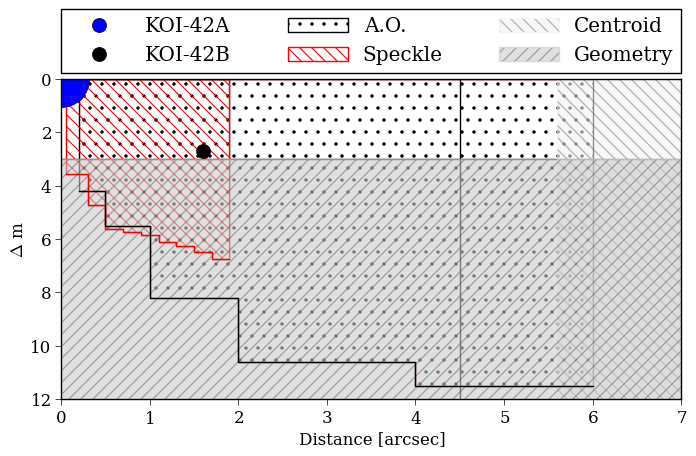}}
\caption{\emph{ \footnotesize Magnitude relative to Kepler-410A plotted versus the distance to the star. Kepler-410B is observed at 1.6$\arcsec$ and $\Delta m$ = 2.7 \citep{adams2012}. A large magnitude difference is excluded because of the transit geometry (\S~\ref{sec:transit_geometry}), while a large angular separation can be ruled out by analyzing the centroid (\S~\ref{sec:centroid}). Finally, ground-based photometry using adaptive optics (A.O.) and Speckle imaging (\S~\ref{sec:ao_speckle}) rules out all but a very small area of the parameter space.}}
\label{fig:confirming_planet}
\end{figure}

We model the planetary transit for various degrees of dilution until the transit fits for the ingress and egress get significantly worse (3$\sigma$ on a $\chi^2$ distribution) and we thereby reject transits occurring at a star more than 3.5 magnitudes fainter than Kepler-410A, and therefore exclude this region of the parameter space. This region is shown as the \textit{geometric limit} in hatched-gray in \fref{fig:confirming_planet}.

\subsubsection{Centroid}
\label{sec:centroid}

\textit{Pixel analysis} during transits can also unmask blends. Transits occurring around a slightly offset blended star would lead to centroid shifts on the {\it Kepler} CCD between in transit and out of transit data. A non detection of such shifts can give an upper limit on the brightness of a potential blend as function of projected distance on the sky.

The \textit{Kepler} team runs elaborate vetting procedures to determine if planetary signatures are caused by blends and centroid shifts are part of this procedure. These procedures are described in detail by \citet{bryson2013}. Kepler-410A, however, is a highly saturated star, which invalidates centroid shift measurements that appear in the Data Validation Report\footnote{http://exoplanetarchive.ipac.caltech.edu/docs/deprecated/KeplerDV.html}. Visual inspection of the difference images \citep[][Section 5]{bryson2013} in the Kepler-410 Data Validation report gives no indication that the transit source is not on the same pixel ($3.98\arcsec$ by $3.98\arcsec$) as Kepler-410A. This analysis is qualitative, however, and does not rule out the companion star. We therefore rely on other evidence given in this paper that the transit occurs on Kepler-410A.

\subsubsection{Ground-based photometry}\label{sec:ao_speckle}
\citet{adams2012} and \citet{howell2011} independently observed a blended object (Kepler-410B) at a distance of $1.6\arcsec$. These observations can also be used to exclude further objects inside certain magnitude and separation limits. The limits from the adaptive optics (A.O.) observations by \cite{adams2012} are shown in \fref{fig:confirming_planet}. The inner spatial limit for detections is at $0.2\arcsec$, where unseen objects up to a contrast of $4.2$ magnitudes in the \textit{Kepler} bandpass are excluded. This increases to $11.5$ magnitudes at $6\arcsec$.

Speckle images of Kepler-410 at $562$ and $692$~nm by \cite{howell2011} provide even tighter spatial constraints (\fref{fig:confirming_planet}), achieving a magnitude contrast of $3.55$ magnitudes between $0.05\arcsec$ and $0.30\arcsec$, with an increasing contrast up to $1.9\arcsec$. The limits in \fref{fig:confirming_planet} are 3$\sigma$ limits for observations at $562$~nm, while those for $692$~nm are about $0.5$ magnitudes tighter for the closest separations.

We further note that the 562 nm detection of Kepler-410B estimates it to be 4.24 magnitudes fainter than Kepler-410A, which, if the same difference holds in the broader \textit{Kepler} band, would place it below our geometric limit of possible planet hosting stars. However, \citet{howell2011} note that at $1.6\arcsec$ separation, the magnitude estimation of detected targets might be underestimated and we choose to adopt the \textit{Kepler} magnitude value for Kepler-410B as claimed in \citet{adams2012}, placing it just above our geometric limit. 

\subsubsection{Spectroscopy}

Spectroscopic observations of Kepler-410 were taken with the HIRES\footnote{High Resolution Echelle Spectrometer on the Keck observatory.} echelle spectrometer at the Keck I telescope and reduced following a procedure described in \cite{chubak2012}. The spectra have a spectral resolution of $R = 55 000$ and stellar lines in the the near-IR wavelength region 654 - 800 nm were used to calculate the Doppler shift. The wavelength scale was determined from thorium-argon lamp spectra taken in twilight before and after each observing night while the wavelength zero-point was determined using telluric lines (from the A and B absorption bands) present in the target spectra. Due to the relatively high $v \sin i$ of the star (see Table \ref{table:spectroscopic_parameters2}), the errors listed here are slightly higher than the typical value (0.1km/s) stated in \cite{chubak2012}. The data are listed in Table~\ref{tab:rv}. We will use these RVs later to constrain scenarios involving binary systems.

\begin{table}
\caption{\emph{ \footnotesize Radial velocity measurements of Kepler-410.}}
\label{tab:rv}
\centering
\begin{tabular}{c c}
\hline\hline\\[-0.35cm]
Date (JD) & Radial Velocity (km/s)\\[0.05cm]
\hline\\[-0.3cm]
2454988.979733  & -40.30 $\pm$ 0.4\\
2455318.048353  & -40.995 $\pm$ 0.3\\
2455726.094382  & -40.18 $\pm$ 0.6\\
\hline
\end{tabular}
\end{table}

\subsubsection{Spitzer observations}
\label{sec:spitzer}

Kepler-410 was observed on 11 July and 18 December 2010 in-transit with the \textit{Spitzer Space telescope} \citep{werner2004}. The first visit consists of full-frame images with a longer integration time and lower accuracy than the second visit, which used \textit{Spitzer}'s subarray mode. We only analyze the subarray data. They consist of 310 sets of 64 individual subarray images, obtained using IRAC's channel 2 \citep{fazio2004}, which is centered at 4.5 $\mu$m. The data are available for download from the \textit{Spitzer} Heritage Archive database\footnote{http://sha.ipac.caltech.edu/applications/Spitzer/SHA} as basic calibrated data (BCD) files. The first observations (which are often more noisy due to the telescope's ramp up) are often ignored \citep[see \eg][]{knutson2008}, but we omit the first 55 observations to keep an equal amount of observations before and after the transit (62 observations on each wing, with 131 in-transit observations).

We analyzed the data following a procedure described by \cite{desert2009}. A square aperture ($11\times11$ pixels) is used to collect the stellar flux (where 64 images of each subarray observation are immediately combined) and the centroid position is calculated. Since a pixel spans 1.2\arcsec, the flux contains the combined light of Kepler-410A and Kepler-410B. Subsequently, a linear function in time is used to de-trend the data, in combination with a quadratic function of the $x$ and $y$ coordinates of the centroids, resulting in five free fitting parameters, \citep[see e.g.][]{knutson2008,desert2009,demory2011} to correct for the pixel-phase effect. We fit only the out-of-transit data, but correct the full dataset.

\begin{figure}[htb]
\centering
\resizebox{\hsize}{!}{\includegraphics{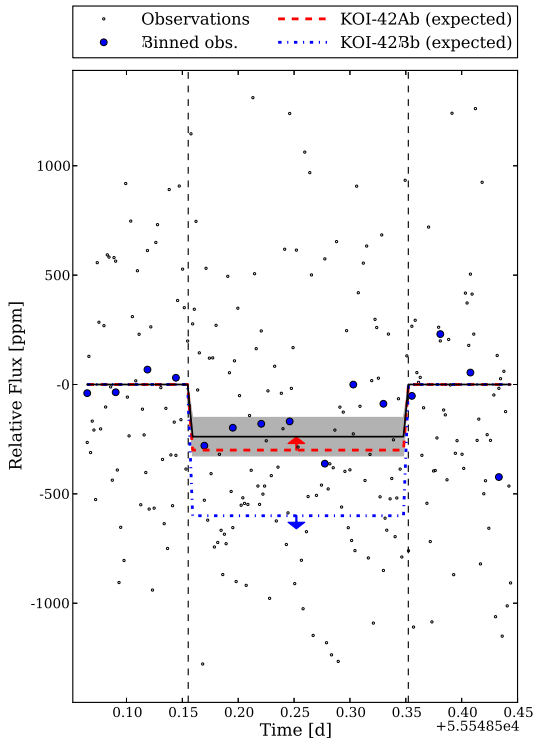}}
\caption{\emph{ \footnotesize The reduced \textit{Spitzer} observations are shown, together with the best fitted transit model (black) and a 1$\sigma$ confidence interval. The blue dots show binned data points. The red dotted line indicates a lower limit for the expected transit depth if the transit occurs on Kepler-410A. The blue dotted line shows a minimum depth if the transit would occur on Kepler-410B.}}
\label{fig:spitzer_transit}
\end{figure}

Now we compare the average flux level of the in-transit data to the out-of-transit data finding a transit depth of $240 \pm 90$~ppm. The uncertainty is calculated by bootstrapping (we re-sample without replacement, treating the in-transit and out-of-transit data separately), which we find to result in a slightly higher error level compared to simply using the scatter on the data points. We adopt this value and show the result in \fref{fig:spitzer_transit}. A similar procedure, comparing median flux levels rather than mean flux levels, gives a transit depth of $260 \pm 90$~ppm.

\subsubsection{Asteroseismology}
\label{sec:asteroseismic_constraints}

Finally, the \textit{Kepler} data provide an asteroseismic constraint on additional objects, by looking at the (absence of) stellar pulsations in the power spectrum (see \fref{fig:power_spectrum}). We searched the power spectrum for excess power from stellar oscillations using the so-called \emph{MWPS} method \citep[see][]{2012MNRAS.427.1784L}. With this, only one set of (solar-like) pulsations was detected, which can be attributed to Kepler-410A because of their high amplitudes, and we can thereby rule out additional signal from bright, large stars to be present in the light curve. We exclude solar-like oscillations of main-sequence stars or red giants up to $K_\textrm{p} = 13$, the geometric exclusion limit (\fref{fig:confirming_planet}).

We can translate this magnitude limit on additional solar-like oscillations into limits on the surface gravity using the method developed by \citet[][]{2011ApJ...732...54C} \citep[see also][]{campante_to_come}.
We estimate a lower limit for the value of $\nu_{\rm max}$\footnote{The frequency at which the oscillations have the largest amplitude.} for a marginal detection of oscillations in the power spectrum. This lower limit on  $\nu_{\rm max}$ can in turn be translated into a lower limit for the surface gravity ($g$) of the star (or $\log\,g$ as most often used) via the simple relation:
\begin{equation}
g \simeq g_{\sun} \left( \frac{\nu_{\rm max}}{\nu_{\rm max, \sun}} \right) \left(\frac{T_{\rm eff}}{T_{\rm eff,\sun}}  \right)^{1/2}.
\end{equation}
The above relation builds on the proportionality between $\nu_{\rm max}$ and the acoustic cut-off frequency \citep[$\nu_{ac}$; see, \eg,][]{1991ApJ...368..599B,2011A&A...530A.142B}. In addition, the procedure uses various scaling relations for \eg\ the amplitudes of the oscillations and the stellar noise background - we refer the reader to \citet[][]{2011ApJ...732...54C} for further details.

For temperatures in the range $T_{\rm eff} = 5500 - 5777$~K we estimate that non-detection of oscillations in any second component (i.e. a star other than Kepler-410A) sets limiting (lower-limit) values for $\log\,g$ of ${\gtrsim} 4.51 \pm 0.05\,\rm dex$ (5500 K) and ${\gtrsim} 4.57 \pm 0.05\,\rm dex$ (5777 K).
For higher assumed values $T_{\rm eff}$, the limiting values for $\log\,g$ are inconsistent with allowed combinations for $\log\,g$ and $T_{\rm eff}$ from stellar evolutionary theory. From these limiting values for $\log\,g$ any potential second component must necessarily be a small dwarf star.

For Kepler-410 the asteroseismic constraint, together with the geometric constraint, is enough to establish the planetary nature of the transit signal. As shown in Section~\ref{sec:transit_geometry} the signal cannot occur on a star fainter than $K_\textrm{p} = 13$ (limiting the maximum true transit depth) and due to the asteroseismic constraint, any object brighter than this is necessarily small. Since the transit depth is given by the size of the transiting object relative to its host star, the two constraints together limit the size of the transiting object to be smaller than Jupiter. For both constraints, observations in a short-cadence sampling are crucial.


\subsection{Scenarios\label{sec:confirmation_scenarios}}

We now use the constraints established in the last section to evaluate three possible scenarios which could cause the transit signal; a chance alignment with a background system (\S~\ref{sec:chance_alignment}), an unseen companion to Kepler-410 (\S~\ref{sec:physical_companion}), and a planet in orbit around Kepler-410B (\S~\ref{sec:koi42b}). Given the available data we can rule them out and conclude that the transit signal occurs on Kepler-410A.


\subsubsection{Chance alignment}
\label{sec:chance_alignment}
The scenario of a background system, largely diluted by a much brighter foreground object (Kepler-410A), is disfavored by a combination of the geometric constraints and the additional observations described in \S~\ref{sec:confirmation_constraints}. With most of the parameter space ruled out, a relevant system would need to have a $K_\textrm{p}$ between $9.5$ and $13$ (see \S~\ref{sec:transit_geometry}) and a separation which is less than $0.02\arcsec$ from Kepler-410A (see \S~\ref{sec:ao_speckle}). 

A detailed analysis on false positive scenarios can be found in \cite{fressin2013}. Following a similar approach we use the Besancon model of the galaxy \citep{robin2003} to simulate the stellar background around Kepler-410. This leads to the prediction of $319$ objects brighter than $13$th magnitude in the R-band (which is close to the \textit{Kepler} band\footnote{\textit{Kepler} magnitudes are nearly equivalent to R band magnitudes \citep[][]{2010ApJ...713L..79K}.}), in an area of one square degree. This places on average $6 \times 10^{-8}$ background stars of sufficient brightness in the confusion region of $0.05$\arcsec around Kepler-410A, the region which is not ruled out by any constraints (see \fref{fig:confirming_planet}). Even without further consideration of whether any background objects could be eclipsing binaries or hosting a transiting planet, we consider this number too small for such a scenario to be feasible. From here on we therefore assume that the transit signal is not caused by a chance 
alignment of a background system.


\subsubsection{Physically associated system}
\label{sec:physical_companion}

We now consider the possibility that the transit occurs on a star physically associated to Kepler-410A but not Kepler-410A itself. According to \citet{fressin2013}, transiting planets on a physically associated star are the most likely source of false positives for small Neptunes. Prior to constraints, they estimate 4.7 $\pm$ 1.0\% of the small Neptune \textit{Kepler} candidates are misidentified in this way. 

For Kepler-410 the spatial constraints from the ground-based photometry (see \S~\ref{sec:ao_speckle}) are far more strict than what was used by \citet{fressin2013}, who only use the \textit{Kepler} data itself to determine the region of confusion.

From the transit geometry stars fainter by $\Delta K_\textrm{p} = 3.5$ are already excluded as possible host stars. Since a physical companion would have the same age as Kepler-410A, we can use the mass-luminosity relation for main sequence stars to derive a lower mass limit. We find this to be about 0.5 M$_\odot$. Furthermore, the companion star cannot be more massive than Kepler-410A itself, otherwise it would be more luminous and thereby visible in the spectra and produce an asteroseismic signal.

We proceed with a simple calculation to quantify the chance that Kepler-410 has an unseen companion with a planet that causes the transit signal. As in \cite{fressin2013}, we assign a binary companion to Kepler-410A following the distribution of binary objects from \cite{raghavan2010}; a random mass ratio and eccentricity and a log-normal distribution for the orbital period. We calculate the semi-major axis using Kepler's third law and assign a random inclination angle, argument of periastron, and orbital phase to the system. 

From the simulated companions, we reject those with a mass lower than 0.5 M$_\odot$. We calculate their angular separation (using the distance estimate from Table~\ref{table:final_parameters}) and reject those which would have been detected in the ground-based photometry. Finally, we compute the radial velocity (RV) signal the companion would produce at the times of the RV measurements (Table \ref{tab:rv}) and reject those objects inconsistent with the observations. For this, we calculate the $\chi^2$ value for each simulated companion, and assign a chance of rejection to each one based on the $\chi^2$ distribution.

We find that only 0.46\% of the simulated objects could pass these tests. The frequency of non-single stars is 44\% \citep{raghavan2010}, resulting in a chance of 0.2\% that an undetected star is associated with Kepler-410A.  This limit would  be even lower if we assume Kepler-410B is physically associated with Kepler-410A, since the probability of additional companions in a multiple system is lower than the value quoted above \citep[an estimated 11\% of all stars are triple system or more complex;][]{raghavan2010}. More elaborate simulations could also further reduce this statistical chance, as we have not taken into account the \textit{Spitzer} transit depth, visibility in spectra, or visibility of asteroseismic features, of this hypothetical companion.


\subsubsection{Kepler-410B}
\label{sec:koi42b}

While the nature of Kepler-410B is largely unknown, some information on the star ca2012ApJn be derived from the observations by \cite{adams2012}. Using their 2MASS $J$ and $Ks$ magnitude, we can convert the measured brightness difference into a temperature estimate, using color-temperature transformations as described by \cite{casagrande2010}. We find a temperature of around 4850 K, assuming a solar metallicity. This indicates a small (dwarf) star, which is consistent with the non-detection of an asteroseismic signal of the object in the blended \textit{Kepler} light (see \S~\ref{sec:asteroseismic_constraints}). 

There is modulation signal present in the \textit{Kepler} data, which is presumably caused by the rotation of Kepler-410B. It indicates a brightness variation of the object of $\approx$ 2.5 \% \citep[assuming the brightness contrast by][see \S~\ref{sec:ao_speckle}]{adams2012}, over a rotation period of 20 days. In fact, the modulation signal has previously been mis-attributed to Kepler-410A \citep{2013arXiv1308.1845M}, resulting in a rotation period inconsistent with what we derive through asteroseismology ($5.25 \pm 0.16$ days, see \S~\ref{sec:splitting}). 

The different colors of Kepler-410A and Kepler-410B can be used to rule out Kepler-410B as a host star, by comparing the transit depth measured in the \textit{Spitzer} IRAC band with the depth as measured by \textit{Kepler}. Kepler-410B is $2.7$ magnitudes fainter than Kepler-410A in the \text{Kepler} band \citep{adams2012}. The flux of Kepler-410B is ${\approx} 8$\% the flux of Kepler-410A. In 2MASS $Ks$ $(2.1 \mu$m) the magnitude difference reduces to 1.9 ($\approx 17$\% flux). We conservatively assume that in \textit{Spitzer}'s IRAC band (4.5 $\mu$m), $\Delta m$ $\leq$ 1.9. Using this assumption, a transit occurring on Kepler-410A would be blended somewhat more in the \textit{Spitzer} observations (depth $\leq 300$ ppm), while a transit occurring on Kepler-410B would only be diluted by less than 
half the dilution in the \textit{Kepler} light (depth $\geq$ 600 ppm).

A measured \textit{Spitzer} transit depth of $240 \pm 90$ ppm distinctly (at a 4$\sigma$ level) rules out Kepler-410B as a potential host star to the transiting planet and is consistent with the planet occurring on Kepler-410A. From here on, we assume that the transits occur on Kepler-410A.


\section{Planetary analysis}\label{sec:planetary_parameters}

\subsection{Period and transit timing variations}
\label{sec:period}

For the planetary analysis, we start from the same dataset as for the asteroseismic analysis (see \S~\ref{sec:asteroseismology}), where we normalize the planetary transits by fitting a second-order polynomial to the transit wings. To determine the planetary parameters we first create a phase folded high signal-to-noise light curve out of the {\it Kepler} light curve. As transit timing variations (TTVs) are present (see below) we cannot simply co-add the light curve on a linear ephemeris but we use the following steps:

\begin{enumerate}
\renewcommand{\theenumi}{\roman{enumi}}
 \item Estimate the planetary period and produce a phase-folded light curve;
 \item Use the phase-folded light curve as an empirical model for the shape of the transit and use this model to determine individual transit times;
 \item Repeat the first two steps until convergence is reached;
 \item Determine TTVs and produce a phase-folded lightcurve which takes this into account;
 \item Model the transit, taking into account the dilution caused by Kepler-410B.
\end{enumerate} 

We find the usage of the phase-folded light curve as an empirical model for the transit quite efficient in determining the times of individual transits. The time for an individual transit event is determined by shifting the empirical model around the predicted transit time. The new time for the transit event is determined by comparing data points with the time-shifted empirical model and minimizing $\chi^2$. Based on the new transit times, a new period estimate can be made and the procedure can be repeated. Following this approach, we reached convergence after only two iterations. 

\begin{figure}[ht]
\resizebox{\hsize}{!}{\includegraphics{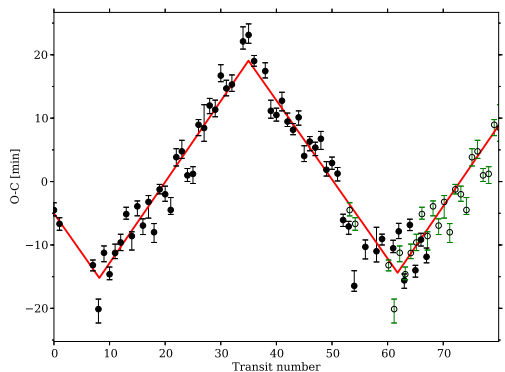}}
\caption{\emph{ \footnotesize O-C diagram showing the observed transit times minus the calculated transit times following a Keplerian orbit (\eqref{eq:linear_period}).  The black points represent individual transit measurements (with their error bars), the green dots are a copy of the observed data points, offset by one full period. They are for illustration only, and were not included in the fit. A clear trend is visible, which is fitted by a model with discontinuities at the turning points.}}
\label{fig:oc_diagram}
\end{figure}

After convergence is reached on determining transit times of individual transit events, the planetary period can be determined. Under the assumption of a perfectly Keplerian orbit, the planetary period is given by a linear interpolation of the transit times:

\begin{equation}
 T (n) = T (0) + n \times \mathrm{Period} \label{eq:linear_period},
\end{equation}
where $T(n)$ and $T(0)$ refer to the $n$th and 0th transit times (taking into account possible data gaps). The period found in this way is $17.833648 \pm 0.000054$ days. Subsequently, we produce an O-C ($Observed-Calculated$) diagram  in which for each transit the calculated transit time is subtracted from the observed transit time, and which we present in \fref{fig:oc_diagram}. Transit Timing Variations (TTVs) are clearly visible. 

The interpretation of TTVs is difficult. Short-period trends can be caused by stellar variability (\eg\ stellar spots causing an apparent TTV signal), while longer-period trends such as here are in most cases attributed to a third body (\eg planet), whose gravitational influence causes the deviation from the strictly Keplarian orbit.

The signal can be highly degenerate, with bodies in or close to different resonance orbits resulting in very similar TTV signals. Attempts of interpretations have been made by performing three-body simulations, with unique solutions for non-transiting objects in only a limited number of cases \citep[see, \eg,][]{nesvorny2013}. TTVs have been successfully used to characterize systems with multiple transiting exoplanets, by studying their mutual gravitational influence \citep[e.g.][]{carter2012}. We have made a visual inspection of the time series to look for additional transit signals, but found none. 

Based on limited data, \citet{ford2011} reported a possible detection of TTVs in the orbit of Kepler-410, and a study of TTVs on the full sample of KOIs \citep{mazeh2013} resulted in an amplitude of $13.95 \pm 0.86$ minutes and a period of $990$ days (no error given) for Kepler-410, using a sinusoidal model. We find a peak-to-peak amplitude of $0.023$ days ($33$ minutes), and a period of $957$ days, not using a sinusoidal but a \textit{zigzag} model, as indicated in \fref{fig:oc_diagram} by the solid line. It is not immediately clear what is causing the seemingly non-sinusoidal shape of the TTVs \citep[see \eg ][for a discussion]{nesvorny2009}. A similar shape is seen for Kepler-36 \citep{carter2012}, where discontinuities occur when the planets are at conjunction. We speculate that the eccentricity of Kepler-410A b could be influencing the shape (see \S~\ref{sec:eccentricity}).

\subsection{Parameters}

\label{sec:parameters}

The period-folded data are then used to determine the planetary parameters. The blending from Kepler-410B (see \S~\ref{sec:ao_speckle}) needs to be taken into account before estimating the planetary parameters, so we subtract the estimated flux due to Kepler-410B (8\%) from the light curve before starting our analysis. 

\begin{table}
\caption{\emph{ \footnotesize Stellar parameters are derived from asteroseismic modeling. Values are from the best fitting model without overshoot; the differences between these values and the ones from the best fitting model including overshoot are taken as a measure of the systematic error from differing input physics in the modeling; this difference is added in quadrature to the uncertainties from the grid optimization. Planetary values are derived from transit modeling combined with asteroseismic results.}}
\label{table:final_parameters}
\begin{center}
\begin{tabular}{l c}
\hline\\[-0.3cm]
Stellar parameters							& {Kepler-410A} \\
\hline\\[-0.35cm]
Mass [M$_\odot$] 							&  	1.214 $\pm$	0.033\\
Radius $R_\star$ [R$_\odot$]						& 	1.352 $\pm$	0.010\\
$\log\,g$ [cgs] 								&  	4.261 $\pm$	0.007\\ 
$\rho$ [g cm$^{-3}$] 							& 	0.693 $\pm$	0.009\\
Age [Gyr] 								& 	2.76  $\pm$	0.54\\
Luminosity [L$_\odot$]					 		& 	2.72 $\pm$	0.18 \\
Distance [pc] 								& 	132 $\pm$	6.9 \\ 
Inclination $i_\star$ [$\rm ^{\circ}$] 					& 	$82.5^{+7.5}_{-2.5}$ \\ 
Rotation period$^*$, $P_{\rm rot}$ [days]				& 	$5.25 \pm 0.16$ \\[0.05cm]
\hline\\[-0.3cm]
Model parameters 							& \\
\hline\\[-0.35cm]
Rotational splitting, $\nu_s$ [$\rm \mu Hz$]				& $2.206^{+0.067}_{-0.065}$ \\
$v\sin(i_\star)^{\dagger}$ [$\rm km s^{-1}$] 				& $12.9 \pm 0.6$ \\ 
$(V_1/V_0)^2$ 								& $1.796^{+0.090}_{-0.085}$\\  
$(V_2/V_0)^2$ 								& $0.861^{+0.073}_{-0.068}$\\[0.05cm] 
\hline\hline\\[-0.3cm]
Planetary parameters 							& Kepler-410A b\\
\hline\\[-0.35cm]
Period [days] 								& 17.833648 $\pm$ 0.000054  	\\
Radius $R_\mathrm{p}$ [R$_\oplus$] 					& 2.838 $\pm$ 0.054	 \\
Semi-major axis $a$ [AU] 						& 0.1226 $\pm$ 0.0047	\\
Eccentricity $e$ 						& 0.17$^{+0.07}_{-0.06}$\\[0.05cm] 
Inclination $i_p$	[$^\circ$]					& 87.72 $^{+0.13}_{-0.15}$ 				\\[0.1cm]
\hline\\[-0.3cm]
Model parameters 							& \\
\hline\\[-0.35cm]
a/R$_*$ 								& 19.50 $^{+0.68}_{-0.77}$ 			\\[0.1cm]
R$_\mathrm{p}$/R$_*$ 								& 0.01923 $^{+0.00034}_{-0.00033}$ 			\\[0.1cm]
Linear LD 								& 0.57 $^{+0.22}_{-0.28}$ 				\\[0.1cm]
Quad LD 									& -0.04 $^{+0.26}_{-0.22}$ 			\\[0.1cm]
\hline
\end{tabular}
\end{center}
$^*${\footnotesize Found as $P_{\rm rot}=1/\nu_s$, and using the uncertainty (asymmetric uncertainties are added quadrature) on $\nu_s$ to find uncertainty for $P_{\rm rot}$.}\\
$^{\dagger}${\footnotesize Found via \eqref{eq:vsini} and using the uncertainties (asymmetric uncertainties are added quadrature) on the parameters $R$, $\nu_s$, and $i$.}
\end{table}

The transits are fitted using the Transit Analysis Package (TAP) which is freely available \citep{gazak2012}. An MCMC analysis is carried out, using the analytical model of \citet{mandel2002}. An orbital eccentricity of zero is assumed for the entire fitting procedure. Flat priors were imposed on the limb darkening coefficients, and they were simply treated as free parameters in our approach. The folded datasets were binned to improve the speed of the MCMC procedure. \fref{fig:transit_fit} shows the transit curve. A list of all parameters is provided in Table \ref{table:final_parameters}.

We finally note that the true errors are likely to be slightly larger than the formal errors reported in Table~\ref{table:final_parameters}. These are the result of the MCMC fitting procedure, and do not take into account systematics in the \textit{Kepler} data \citep{vaneylen2013}, or the uncertainty in the flux contribution by the blended light from Kepler-410B, both of which could affect the transit depth.

\subsection{Planetary eccentricity}
\label{sec:eccentricity}

We have access to two estimates of the stellar density. One value was obtained from the asteroseismic modeling of the stellar pulsations ($\rho_\textrm{asteros.}$) and one from modeling the planetary transit \citep[${\rho_\textrm{transit}}$;][]{seager2003,tingley2011},

\begin{equation}\label{eq:stellar_density}
 \rho_\textrm{transit} = \frac{3 \pi}{G P^2} \left( \frac{a}{R_*}\right)^3 = 0.441 \pm 0.050 \mathrm{~g/cm}^3 \, ,
\end{equation}
where $G$ is the gravitational constant and all other parameters are listed in Table~\ref{table:final_parameters}. To obtain an estimate of $\rho_\textrm{transit}$ a particular orbital eccentricity ($e$) needs to be assumed, which in this equation was set to zero. Therefore calculating the ratio of the two density estimates leads to a lower limit on the orbital eccentricity. 

\begin{figure}[ht]
\centering
\resizebox{\hsize}{!}{\includegraphics{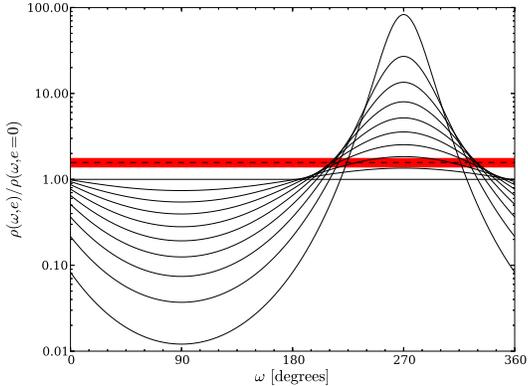}}
\caption{\emph{ \footnotesize Change of perceived stellar density, compared to zero-eccentricity density, for different angles of periastron. The inner solid line represents an eccentricity of $e=0$, the outer depicts $e=0.9$. The dashed line gives the location of the density ratio given in \eqref{eq:eccentricity_periastron} for Kepler-410 (uncertainty of the ratio is given by the red band).}}
\label{fig:eccentricity}
\end{figure}

Following the notation in \cite{dawson2012} we obtain

\begin{equation}
\label{eq:eccentricity_periastron}
\frac{\rho_\textrm{asteros.}}{\rho_\textrm{transit}} = \frac{(1-e^2)^{3/2}}{(1+e\sin \omega)^3} = 1.57 \pm 0.18 \, ,
\end{equation}
where $\omega$ is the argument of periastron and we took $\rho_\textrm{asteros.} = 0.693 \pm	0.009$ from Table~\ref{table:final_parameters}. As the value is not consistent with unity within error bars, a circular orbit for the planet is ruled out. The eccentricity is a function of $\omega$, as can be seen in \fref{fig:eccentricity}.

\begin{figure}[ht]
\centering
\resizebox{\hsize}{!}{\includegraphics{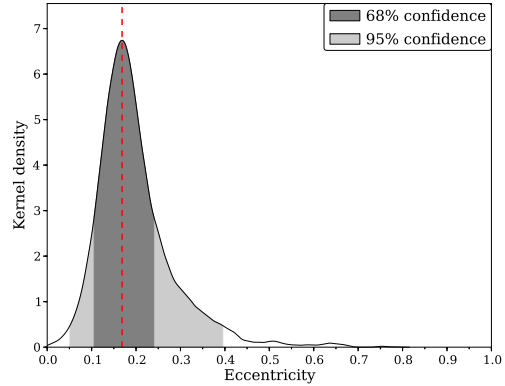}}
\caption{\emph{ \footnotesize Kernel density distribution of the eccentricity values. The mode (dotted red line) is seen at an eccentricity of 0.17 and the uncertainties (highest posterior density credible regions) are indicated in grey.}}
\label{fig:eccentricity_pdf}
\end{figure}

\fref{fig:eccentricity} and \eqref{eq:eccentricity_periastron} indicate that a lower limit on the system's eccentricity can be derived. For certain arguments of periastron (around $\omega \approx 210^\circ$ or  $\omega \approx 320^\circ$), high eccentricities cannot be excluded. However, the range of periastron angles becomes increasingly narrow for increasing eccentricities. Taking a sample assuming random angles of periastron, and a Gaussian distribution for $\rho_\textrm{asteros.}/\rho_\textrm{transit}$ to take into account the uncertainty of \eqref{eq:eccentricity_periastron}, and using a correction factor for non-grazing transits as described in \cite{dawson2012}, we find that the mode of the eccentricity is $0.17$ and 68\% of the eccentricities are contained in the interval [0.11,0.24], as indicated on \fref{fig:eccentricity_pdf}.

For the above analysis to deliver unbiased eccentricity results, it is important to remove TTVs (see \S~\ref{sec:period}) and third light from Kepler-410B (\S~\ref{sec:parameters}) from the light curve \citep[see \eg][]{kipping2013}. We expect no additional light dilution because of the additional constraints presented in \S~\ref{sec:constraints}, and specifically the asteroseismic constraint in \S~\ref{sec:asteroseismic_constraints}  which rules out bright companion stars.


\subsection{Stellar obliquity}
\label{sec:obliquity}

The stellar obliquity (the opening angle between the stellar rotation angle and the orbital angular momentum) can be constrained with our asteroseismic modeling and transit measurements. We measure  similar values for the inclination of the stellar rotation axis ($i_\star=82.5^{+7.5}_{-2.5}$ [$^\circ$]) and the planetary orbital axis ($i_p=87.72\pm0.15^\circ$). We have no information on the other variable defining the stellar obliquity, its projection on the plane of the sky. However we can ask how likely it would be that we measure similar inclinations if the orientation of the stellar rotation axis is uncorrelated to the planetary orbit and randomly oriented. For this we look at a distribution which is flat in $\cos i_\star$, and which leads to a random orientation of the angular momentum axis on a sphere. This way we find that there there is  a 17\% chance to find the stellar rotation axis inclined as close to 90$^\circ$ as is the case, assuming no correlation between the stellar rotation axis and the 
planetary orbital inclination. Therefore our asteroseismic measurement of $i_\star$ suggests a low obliquity in the Kepler-410A system. 

\section{Discussion}
\label{sec:discussion}

With the validation of Kepler-410b as a small Neptune-sized exoplanet ($2.838 \pm 0.054$~R$_\oplus$), it joins the current list of 167 confirmed {\it Kepler} exoplanets around 90 stars. Thanks to the sampling in short cadence and the brightness of the host star, the stellar (and therefore planetary) parameters are known to very high accuracy. The star is the third brightest of the current sample of confirmed \textit{Kepler} planet host stars, only preceded by Kepler-21 \citep{howell2012} and Kepler-3 \citep[HAT-P-11][]{bakos2010}, both of which have a period of only a few days. 

Even outside the \textit{Kepler} field, only about 10-20 planets (which are typically not as well-characterised) are known to transit around stars which are brighter or of similar brightness. 55 Cnc e \citep{mcarthur2004} is the brightest and together with HD 97658 \citep{howard2011,dragomir2013} and Kepler-21b \citep{howell2012}, they are the only planets smaller than Kepler-410A b around stars brighter than Kepler-410A, and they all have a shorter orbital period. The only ones with longer orbital periods are the Jupiter-sized planets HD 17156b \citep{fischer2007} and HD 80606b \citep{naef2001}. Perhaps the most similar system is the bright star Kepler-37, which has three planets of sub- and super-Earth size on orbital periods of 13, 21 and 39 days \citep{barclay2013}.

That the host star can be well-studied has its implications on the planetary parameters, which are now also well-known. This makes Kepler-410 an interesting object for follow-up observations. High-quality radial velocity observations might be able to constrain the planetary mass and therefore also its density, and, if the latter is favorable, even transmission spectroscopy might be within reach of some instruments. In addition, such observations might shed more light on the observed transit timing variations, which we suspect are caused by one or more additional planets in the system. With a relatively high TTV amplitude ($\sim$ 30 minutes peak-to-peak), one might hope (an) additional planet(s) can be revealed with radial velocity observations. Full simulations of the observed TTVs were beyond the scope of this paper but might be fruitful due to the eccentricity of Kepler-410A b; our observed transit times are available upon request.

Our finding of a low obliquity in Kepler-410A can be compared to obliquity measurements in other exoplanet systems with multiple planets. The first multiple system for which the projected obliquities has been measured is the Kepler-30 system which harbors three transiting planets \citep{sanchisojeda2012}. The authors found a low projected obliquity by analyzing spot crossing events. \cite{2013ApJ...766..101C} found good alignment between the orbital and stellar inclinations for Kepler-50 and Kepler-65, analyzing the splitting of the rotational modes in a similar way as presented in this work. \cite{hirano2012} and \cite{albrecht2013} analyzed the KOI-94 and Kepler-25 systems and found low projected obliquity. For the multiple transiting planet system KOI-56, asteroseismic modeling revealed a high obliquity between the orbit of the two planets and the stellar rotation \citep{huber2013}. The authors suggest that a companion leads to a misalignmemt of one planet, which then influenced the orbital plane of the 
other planet. With the measurements at hand, it appears as if the obliquity distribution for multiple planet systems is flatter than what is observed for systems with single close-in Jupiter-sized planets \citep{albrecht2012,albrecht2013}. 

One way to learn more about the obliquity in Kepler-410A would be to also measure the projection of the stellar rotation axis via the Rossiter-McLaughlin (RM) effect \citep[][]{1924ApJ....60...15R,1924ApJ....60...22M}. The amplitude of the RM effect would be of the order of a few m\,s$^{-1}$, despite the small transit depth, as $v \sin i_\star$ is large.

Of particular interest in this regard is also the tight constraints derived on the planetary eccentricity, which is measured to be inconsistent with a circular orbit. While eccentricities routinely result from radial velocity observations of exoplanet hosts, this is one of the first stars for which the eccentricity is tightly constrained using only photometric measurements, which to our knowledge has only resulted in excluding circular orbits in the case of Kepler-63b \citep{sanchisojeda2013}.


\section{Conclusions}
\label{sec:conclusion}

Using a combination of high-quality \textit{Kepler} data and ground-based photometry and spectroscopy, we are able to validate the presence of a planet around Kepler-410A; a small Neptune in an orbit with a period of $17.8$ days. This makes Kepler-410A the third brightest \textit{Kepler} planet host star currently known. A detailed analysis of the solar-like oscillations allows for a characterization of the stellar mass to within 3\%, while the radius is known to less than a 1\% and the age is determined to within 20\%. The asteroseismic study also allowed a precise determination of the distance to the star. 

Furthermore, we constrain the rotation rate and inclination angle of the host star and find the results to be consistent with low obliquity. This is a result similar to most obliquity measurements in multiple planet systems, which is in contrast to measurements of obliquities in Hot-Jupiter systems, where the obliquities are much more diverse. With an accurate determination of the stellar density through asteroseismology, we are able to photometrically constrain the planetary eccentricity to 0.17$^{+0.07}_{-0.06}$. We finally note that transit timing variations strongly suggest the presence of at least one additional (non-transiting) planet in the system.\\


\vspace*{1cm}
\small{\emph{Acknowledgements}. We thank Joanna Molenda-\.{Z}akowicz, Lars A. Buchhave and Christoffer Karoff for sharing stellar spectra. We thank Luca Casagrande for help with the InfraRed Flux Method to obtain the stellar distance and David Kipping for helpful comments in reviewing the manuscript. The referee's helpful comments and suggestions have led to significant improvements. MNL would like to thank Dennis Stello and his colleagues at the Sydney Institute for Astronomy (SIfA) for their hospitality during a stay where some of the presented work was done. WJC and TLC acknowledge the support of the UK Science and Technology Facilities Council (STFC). Funding for the Stellar Astrophysics Centre is provided by The Danish National Research Foundation (Grant agreement no.: DNRF106). The research is supported by the ASTERISK project (ASTERoseismic Investigations with SONG and Kepler) funded by the European Research Council (Grant agreement no.: 267864). Funding for the \textit{Kepler} Discovery mission is 
provided by NASA’s Science Mission Directorate. The Spitzer Space Telescope is operated by the Jet Propulsion Laboratory, California Institute of 
Technology under a contract with NASA. We thank the entire \textit{Kepler} team, without whom these results would not be possible. }

\appendix
\section{Model correction from rotational second-order effects}\label{append}
\begin{figure}[ht]
\centering
\includegraphics[scale=0.4]{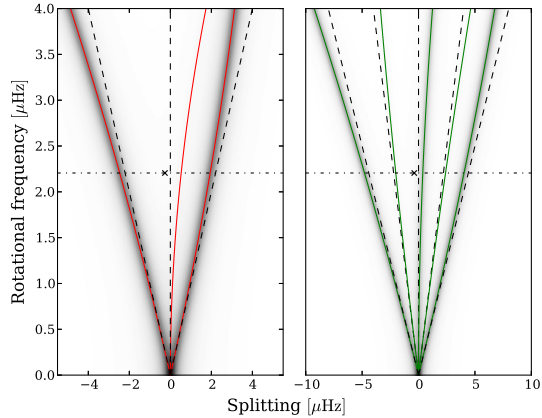}
\caption{\emph{ \footnotesize Illustration of the effect of second order splittings as a function of rotational frequency for an $\ell=1$ mode (left) and an $\ell=2$ mode (right), both with a resonance frequency of $\nu_{n\ell}=2000\rm\,\mu Hz$. Dashed black lines (one for each $m$-component) give the first-order splitting, while the solid lines give the second-order splittings. The gray scale indicates the relative height of azimuthal components when assuming an inclination of $i=82.5^{\circ}$. The dash-dotted horizontal line indicates the obtained frequency splitting from peak-bagging. The adopted mode line width is $\Gamma_{n\ell}=1\rm\,\mu Hz$. The cross gives the mid-point between the $m$-components of $\pm \ell$.}}
\label{fig:second_order}
\end{figure}
We approximate the second-order effect on the rotational splitting by \citep[see][]{1998ESASP.418..385K}:
\begin{eqnarray}
\delta\nu_2 = \left(\frac{\nu_s^2}{\nu_{n\ell}} \right)(\Delta_{n\ell}^{(1)} + m^2\Delta_{n\ell}^{(2)}) 
+ \left(\frac{2\pi\nu_s}{\Omega_{\rm K}} \right)^2\nu_{n\ell} \Delta_{n\ell}^{(3)} Q_{2\ell m}. 
\label{eq:split2}
\end{eqnarray}
Parameters are estimated as in \citet[][]{2010aste.book.....A}, with the exception of $\Delta_{n\ell}^{(3)}$, which is assigned a value of $2/3$ as in \citet[][]{1998ESASP.418..385K}.

The effect of the second-order splittings can be seen in \fref{fig:second_order}. The difference between first-order (dashed) and second-order (solid) splitting increases with the rotational frequency. The gray scale in this figure indicates the relative heights of the $m$-components for the obtained inclination angle of $i=82.5^{\circ}$. An important aspect to notice here is that the sectoral components ($|m|=\ell$) dominate the split multiplet, especially for $\ell=1$. This means that the fitted components will have a separation of $2\ell \nu_s$, whereby the splitting obtained from the peak-bagging will be biased only very little by second-order effects, and we can readily adopt this as the "true" splitting \citep[see also][]{2010AN....331..933B}. However, even though the splitting is little affected by the second-order effects it is clear from \fref{fig:second_order} that the mid point (indicated by the cross) between the fitted $m$-components of value $\pm \ell$ will deviate from the first-order 
rotationally unaffected estimate of the $m=0$ resonance frequency - which is the frequency entering into the stellar modeling. For the case shown in \fref{fig:second_order} where $\nu_{n\ell}=2000\rm\,\mu Hz$ we get from the obtained splitting (combined with computed stellar model) that the $m=0$ frequency for $\ell=1$ is estimated too low by a value of ${\sim}0.26\rm\,\mu Hz$ while $m=0$, $\ell=2$ is estimated too low by a value of ${\sim}0.38\rm\,\mu Hz$. This offset will naturally have some impact on the model computed from the fitted frequency, and in turn the parameters of the model (via $\Omega_{\rm K}$) will affect the estimated impact of second-order effects. 

For the particular case of Kepler-410 the addition of second-order effects on the splittings did not impact the modeled parameters in any significant way, with changes generally on the per mil scale. The stellar age was most affected with a change of ${\sim}1.5\%$. However, we note that even though the effect from the second-order splitting was small for Kepler-410 it might not be for other targets, and it should in general be considered. Estimates of rotational splittings will also generally be biased from the second-order effect in stars with lower inclinations, and an approach as taken here might not be appropriate. For Kepler-410 we can add the effect \emph{a posteriori} as the near equator-on orientation of the star allows for a relatively un-biased estimation of the rotation rate.

\bibliographystyle{bibstyle}
\bibliography{references}

\end{document}